\documentclass[prl,twocolumn,preprintnumbers,amsmath,amssymb,superscriptaddress]{revtex4-2} 

\usepackage[colorlinks,bookmarks=false,citecolor=blue,linkcolor=red,urlcolor=blue]{hyperref}

\usepackage{amsmath,mathrsfs,amsbsy,color,graphicx,bm,amsthm,amsfonts}
\usepackage{bbm}
\usepackage{bm}
\usepackage{times}
\usepackage{units}
\usepackage{dcolumn}
\usepackage{graphicx}
\usepackage{epsfig}
\usepackage{epstopdf}
\DeclareMathSymbol{\shortminus}{\mathbin}{AMSa}{"39}
\usepackage{mathrsfs}
\usepackage{braket}
\usepackage{amssymb}
\usepackage{txfonts}
\usepackage{enumitem}
\usepackage{multirow}
\graphicspath{{fihures/},{pics/}} \usepackage{subfigure} 
\usepackage{microtype}
\usepackage[cmyk]{xcolor} 
\usepackage{placeins}

\usepackage{orcidlink}

\begin{document}
\title{Escape of quantum information across an analogue black hole horizon}
\author{Zhilong Liu~\orcidlink{0009-0000-0353-5113}}
\affiliation{Department of Physics, Key Laboratory of Low Dimensional Quantum Structures and Quantum Control of Ministry of Education, Institute of Interdisciplinary Studies,  Hunan Research Center of the Basic Discipline for Quantum Effects and Quantum Technologies,  and Synergetic Innovation Center for Quantum Effects and Applications, Hunan Normal
	University, Changsha, Hunan 410081, P. R. China}
	
\author{Wentao Liu~\orcidlink{0009-0008-9257-8155}}
\affiliation{Department of Physics, Key Laboratory of Low Dimensional Quantum Structures and Quantum Control of Ministry of Education, and Synergetic Innovation Center for Quantum Effects and Applications, Hunan Normal
	University, Changsha, Hunan 410081, P. R. China}

\author{Zehua Tian~\orcidlink{0009-0004-6581-9890}}
\email{tzh@hznu.edu.cn (Corresponding author)} 
\affiliation{School of Physics, Hangzhou Normal University, Hangzhou, Zhejiang 311121, China}
	
\author{Jieci Wang~\orcidlink{0000-0001-5072-3096}}
\email{jcwang@hunnu.edu.cn (Corresponding author)}
\affiliation{Department of Physics, Key Laboratory of Low Dimensional Quantum Structures and Quantum Control of Ministry of Education, Institute of Interdisciplinary Studies,  Hunan Research Center of the Basic Discipline for Quantum Effects and Quantum Technologies,  and Synergetic Innovation Center for Quantum Effects and Applications, Hunan Normal
	University, Changsha, Hunan 410081, P. R. China}\maketitle

{\bf 
The complete evaporation of black holes, as a natural endpoint of Hawking radiation, gives rise to the black hole information paradox, which fundamentally challenges the principles of unitarity and information conservation in quantum mechanics.  Although the AdS/CFT correspondence indicates that information is preserved during black hole evaporation, the precise mechanism by which it is recovered from the Hawking radiation remains an open question. To explore a potential resolution, we investigate information transfer in a spin-chain model featuring an effective horizon realized via position-dependent couplings within an XY spin chain. We derive and demonstrate Page curve-like behavior, and analyze the transmission of quantum resources, such as entanglement and coherence, across the effective horizon. Our results show that quantum resources initially localized within an interior subsystem can be transferred to the exterior via particle radiation through the horizon. This study provides a novel perspective from quantum simulation on how information may escape from black holes, thereby contributing to the further understanding of the black hole information paradox.}

{\bf Introduction}\\
	Black holes, as extreme astrophysical objects predicted by general relativity, are characterized by the fact that external observers can only detect their total mass, electric charge, and angular momentum, with no access to the internal state of the black hole—a property known as the no-hair theorem. Furthermore, due to quantum effects,  they emit Hawking radiation ~\cite{Hawking:1974rv}, a thermal spectrum whose temperature decreases with increasing black hole mass. As a consequence, in its final stages, a black hole evaporates completely via thermal radiation, seemingly erasing all the information it once contained. This conclusion directly challenges the fundamental principles of information conservation and unitary evolution, giving rise to the black hole information paradox~\cite{Hawking:1976ra}. 
	
	In the context of modern advances in quantum foundations and quantum technologies, abandoning unitarity to resolve this paradox would be a radical and unattractive alternative. Moreover, the AdS/CFT correspondence formulated by Maldacena~\cite{Maldacena:1997re} intimates a unitary holographic description of quantum gravity. Similarly, Hawking~\cite{Hawking:2005kf} has argued that elementary quantum gravitational interactions should neither dissipate information nor destroy coherence, thereby directly challenging the information loss hypothesis. Despite these compelling arguments, the question of how the paradox is concretely resolved remains an active and vigorously debated topic. Numerous studies have since attempted to resolve the paradox from various perspectives, including modifications to the thermal spectrum, the island rule, soft hair, remnants, among others~\cite{Almheiri:2019psf,Penington:2019kki,Qi:2021sxb,Hsin:2020mfa,Huber:2025aah,Alkac:2025hrv,Kraus:1994by,Parikh:1999mf,Chen:2014jwq,Hawking:2016msc}.
	\begin{figure}[t]
		\centering{\includegraphics[width=0.98\linewidth]{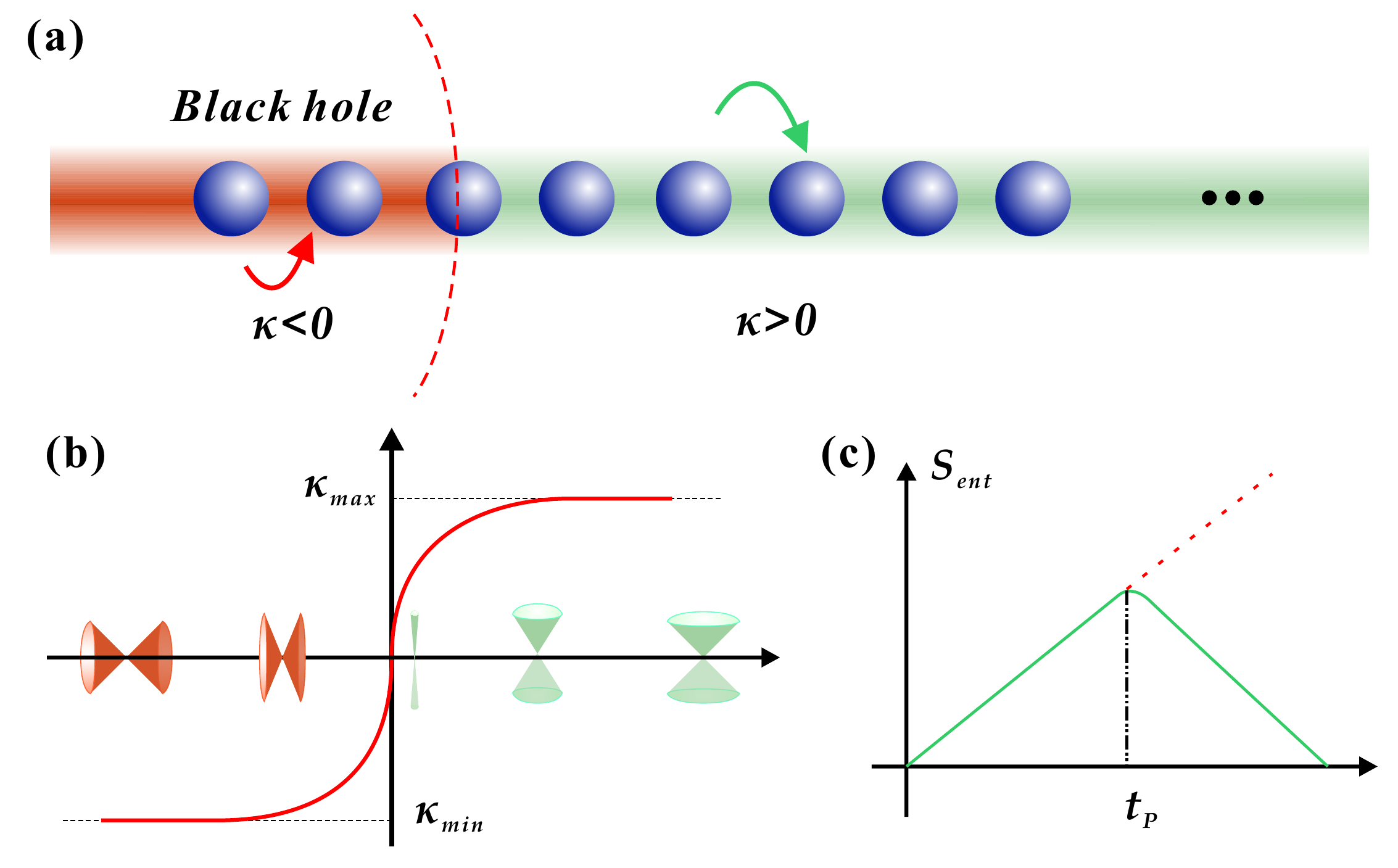}
			\caption{(a) Schematic of the an effective horizon, realized via an XY spin chain with site-dependent couplings $\kappa_n$. The coupling $\kappa$ undergoes a sign reversal between the interior and exterior regions, demarcating the effective horizon. (b) Cross-section of the $\kappa$ distribution and associated light-cone schematic for the one-dimensional black hole model employed herein. (c) Time evolution of the entanglement entropy $S_{ent}$ between the black hole interior and exterior during continuous particle radiation from the interior. The red dashed line represents the semiclassical Hawking radiation prediction, characterized by linear growth; the green solid line depicts the Page curve, which coincides with the semiclassical result in the early regime but, following the Page time $t_P$ (when roughly half the particles have leaked out), exhibits a downturn and monotonic decay to zero upon complete black hole evaporation. Throughout all numerical simulations, we consistently employ a total system size of $L = M + N = 10$ qubits, where the system subspace comprises $M = 2$ qubits and the bath consists of $N = 8$ qubits.}\label{fig_diagram_xy}
		}
	\end{figure}
	
	 From the perspective of a statistical model of black holes, Page computed the entanglement entropy and information between interior and exterior subsystems under the premise of unitary evolution~\cite{Page:1993wv}. It has been demonstrated that the entanglement entropy between these subsystems increases linearly in the early stages of Hawking radiation. However, once the black hole has decayed to approximately half of its original mass—a point often referred to as the Page time—the entanglement entropy begins to decrease, eventually vanishing as the black hole fully evaporates. Concomitantly, information is expected to emerge rapidly after the Page time. In recent years, quantum properties such as quantum entanglement and quantum coherence have garnered substantial research interest across diverse disciplines, such as condensed matter physics, quantum information science, and quantum gravity theory~\cite{Kish:2024rhw,Capizzi:2022igy,Martirosyan:2024rxm,Simon:2011eeg,Bianchi:2021aui,Bianchi:2021lnp,Lydzba:2020qfx}. The study of quantum information dynamics in analogous many body spin chains has likewise attracted considerable attention. Early research showed that in ergodic systems, the entanglement entropy grows linearly in time until saturating at a value dictated by the volume law of excited states~\cite{Calabrese:2005in}. Notably, Kehrein~\cite{Kehrein:2023yeu} elucidated that an analytically solvable system plus bath model for fermionic many body systems manifests entanglement dynamics in accord with the Page curve, identifying a salient correlation between the inflection in entanglement entropy and the particle current, marking the Page time. Furthermore,  Glatthard~\cite{Glatthard:2024pyt} generalized this Page curve dynamics to generic open quantum systems weakly coupled to a low-temperature environment. The endeavor to demonstrate and investigate Page curve like dynamics across diverse quantum systems has catalyzed a surge of research interest~\cite{Ptaszynski:2023edi,Saha:2024ims,Glatthard:2025mbb,Ganguly:2025kil,Li:2025zls,Jha:2025eth,Ray:2025dlf}. 
	 
	 To address the black hole information paradox, a fundamental challenge lies in the experimental observation of Page curve dynamics, which serves as a definitive signature of information unitarity. Meanwhile, theoretical and experimental research on simulating quantum fields in a curved spacetime background under experimentally feasible systems has become increasingly mature~\cite{Kinoshita:2024ahu,Konye:2022wds,Horner:2022sei,Kinoshita:2025fij,Sheng:2018jmy,Vocke:2018xwb,Sharan:2025rsg,Wang:2020ypl,Svancara:2023yrf,Viermann:2022wgw,Shi:2021nkx}. Here, we investigate the information dynamics of a spin-chain system featuring an effective horizon realized via an XY spin chain with spatially varying couplings~\cite{Shi:2023aoz,Yang:2019kbb,Shi:2021nkx}. We provide a comprehensive numerical analysis of the entanglement entropy's evolution, alongside the transport of quantum coherence and entanglement across the horizon. Notably, our simulation parameters are carefully calibrated to align with existing experimental platforms, such as superconducting circuits, ensuring the immediate relevance of our findings to near-term laboratory demonstrations.
	 This approach may help clarify how information could  escape a black hole, thereby contributing to resolving the information paradox.
	
{\bf Models}

We consider a XY spin chain with site-dependent coupling (see Fig.~\ref{fig_diagram_xy}), described by the Hamiltonian
	\begin{equation}
		H_{XY}=\frac{1}{2}\sum_{n=1}^{L-1} \left[-\kappa_n(\sigma^x_n\sigma^x_{n+1}+\sigma^y_n\sigma^y_{n+1})\right]-\sum_{n=1}^{L}\mu\sigma_n^z\label{eq_XY},
	\end{equation}
	where \( \kappa_n \) denotes the nearest-neighbor XY coupling strength, which varies with the lattice site \( n \), and \( \mu \) represents the on-site potential. This type of interaction can be used to simulate quantum field equations in curved spacetime~\cite{Yang:2019kbb,Liu:2024wqj,liu2025scrambling}. For a simple \((1+1)\)-dimensional spherically symmetric black hole model, the line element (with signature \((-, +)\)) is given by
	\begin{equation}
		\mathrm{d}s^2 = -f(x) \mathrm{d}t^2 + \frac{1}{f(x)} \mathrm{d}x^2,
	\end{equation}
	where \( f(x) \) is a function encoding the curvature of the spacetime. By employing variable substitution and spatial discretization, the quantum field equation in this curved background can be mapped to the Heisenberg equation governed by the XY-chain Hamiltonian in Eq. (1). Under this mapping, the site-dependent nearest-neighbor hopping strengths are defined as~\cite{Yang:2019kbb,Liu:2024wqj}
	
	\begin{equation}
		\kappa_n = \frac{f\left[\left(n - \frac{1}{2}\right)d\right]}{4d}.
	\end{equation}	
	Here, \( f(n d) \) corresponds to the spatially discretized form of \( f(x) \), and \( d \) is the lattice spacing. Since the metric function \( f(x) \) changes sign across the horizon, \( \kappa_n \) takes negative values inside the black hole and positive values outside. 
	To further elucidate the dynamical properties of this model, we decompose the Hamiltonian in Eq.~(\ref{eq_XY}) into a standard system–bath form:
	\begin{equation}
		\begin{aligned}
			H &= H_S + H_B + H_I, \\
			H_S &= \frac{1}{2}\sum_{n=1}^{n_h-2} \left[-\kappa_n \left( \sigma_n^x \sigma_{n+1}^x + \sigma_n^y \sigma_{n+1}^y \right) \right] - \sum_{n=1}^{n_h-1} \mu \sigma_n^z, \\
			H_B &= \frac{1}{2}\sum_{n=n_h}^{L-1} \left[-\kappa_n \left( \sigma_n^x \sigma_{n+1}^x + \sigma_n^y \sigma_{n+1}^y \right) \right] - \sum_{n=n_h}^{L} \mu \sigma_n^z, \\
			H_I &= -\frac{1}{2}\kappa_c \left( \sigma_{n_h-1}^x \sigma_{n_h}^x + \sigma_{n_h-1}^y \sigma_{n_h}^y \right),
		\end{aligned}
	\end{equation}
	where the subscripts $S$, $B$ and $I$ stand for system, bath and interaction, respectively. \( n_h \) labels the lattice site corresponding to the black hole horizon. In a physical black hole spacetime, the narrowing of the light cone near the event horizon is reflected in the vanishing of the metric function $f(x)$. In the inhomogeneous spin chain system, this corresponds to a significant suppression of the coupling strength $\kappa_c$ in the vicinity of the horizon, effectively placing the system in the weak-coupling regime. Furthermore, by initializing the exterior region in an empty state, it functions as a low-temperature bath. These features satisfy the criteria established by Glatthard~\cite{Glatthard:2024pyt}, wherein the interplay between a weak system-bath interaction and a vacuum-like environment facilitates the emergence of Page curve like entanglement dynamics.
	
	Assuming the composite system is initially in a pure state, its global purity is preserved under unitary evolution. For a bipartition of the model into subsystems $S$ and $B$, the bipartite correlations are quantified by the von Neumann entanglement entropy
	\begin{equation}
		S_{\text{ent}} = -\operatorname{tr}(\rho_S \ln \rho_S),
	\end{equation}
	where 
	\begin{equation}
		\rho_S = \operatorname{tr}_B(\rho_{\text{tot}}),
	\end{equation}
	is the reduced density matrix of subsystem $S$, and $\rho_{\text{tot}}$ represents the density operator of the total system. In typical quantum many-body systems characterized by local interactions, the entanglement entropy of a subsystem in the ground state generally obeys the area law \cite{Eisert:2008ur}, reflecting the short-range nature of the underlying correlations.
	\begin{figure}[t]
		\centering{\includegraphics[width=0.96\linewidth]{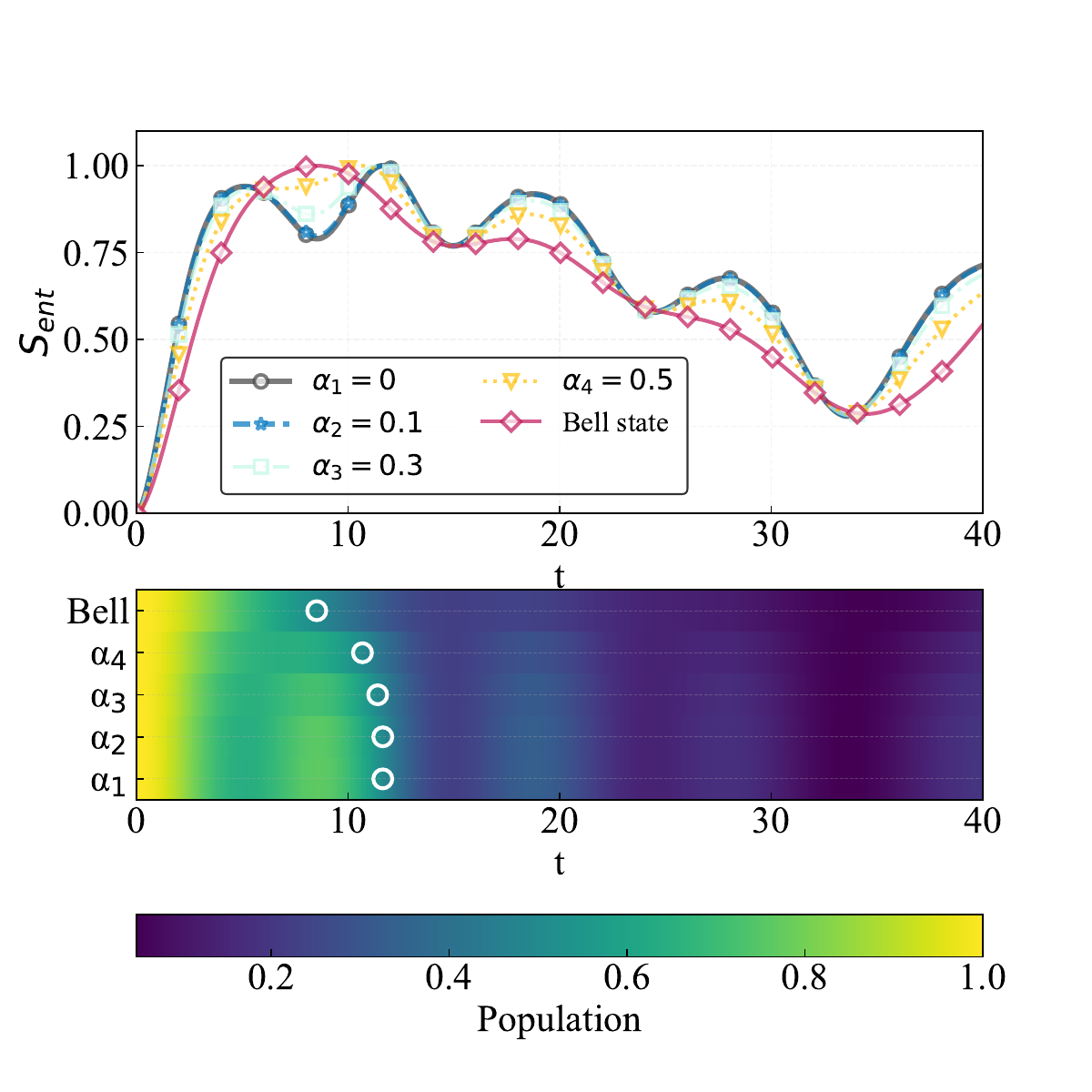}
			\caption{Upper panel: Evolution of the entanglement entropy $S_{ent}$ between the system and the bath in inhomogeneous spin chain system. The different curves correspond to initial states prepared with varying degrees of entanglement inside the system. Lower panel: Time evolution of the particle number inside the system for the same set of initial states. The white circles mark, for each case, the time at which the particle number reaches half of its initial value. The numerical data are obtained for a chain of length \(L=10\) and a discretization interval \(d=2\).}\label{fig_pagecur}
		}
	\end{figure}
	Furthermore, in non-equilibrium ergodic systems, the entanglement entropy generally grows linearly until it saturates to a value determined by the volume law of excited states~\cite{Calabrese:2005in}. This growth aligns with Hawking’s semiclassical derivation, wherein the production of entangled pairs at the horizon—with one particle escaping to infinity while its partner falls into the interior—results in a monotonic increase of the entanglement entropy between the black hole and its environment. However, to preserve global unitarity, Page’s analysis necessitates that this entropy must eventually decrease after the characteristic ``Page time'' ultimately vanishing upon complete evaporation.

	{\bf Results}
	
	\paragraph{{\it Page curve like dynamics}.} In this section, we demonstrate the emergence of Page curve like dynamics within the framework of the Hamiltonian defined in Eq.~(\ref{eq_XY}). We consider an $(1+1)$-dimensional asymptotically flat black hole spacetime, characterized by the metric function 
	\begin{equation}
		f(x) = \beta \tanh(x).
	\end{equation}
	This choice of $f(x)$ ensures a well-defined horizon at the origin and recovers a flat Minkowski geometry in the asymptotic limit $|x| \to \infty$. Within this system-bath framework, the exterior is initialized in the empty state. To systematically compare the evolutionary behavior of entanglement entropy across distinct initial interior configurations, the interior system is prepared in entangled states of varying degrees of entanglement. The overall initial state can thus be expressed as:	
	\begin{eqnarray}
		|\psi\rangle = \left( \alpha|\mathrm{\uparrow \downarrow}\rangle + \sqrt{1 - \alpha^2}|\mathrm{\downarrow \uparrow}\rangle \right) \otimes |\mathrm{\downarrow\downarrow\downarrow} \dots \rangle,\label{eq_phi0ent}
	\end{eqnarray}
	where the parameter \( \alpha \) quantifies the initial entanglement inside the system. When \( \alpha = 0 \), the internal state is a simple single-excitation product state; when \( \alpha = 1/\sqrt{2} \), the internal pair forms a maximally entangled Bell state.
	
	We then evaluate the temporal evolution of the entanglement entropy between the system and the bath, alongside the local occupation numbers inside the system. As shown in the upper panel of Fig.~\ref{fig_pagecur}, different initial states share a similar downward trend following a nearly linear growth of entanglement entropy driven by the progressive leakage of excitations into the bath. After a certain time, however, the system-bath entanglement entropy begins to turn downward and decay. This downward trend persists as the remaining excitations escape, and the entropy asymptotically approaches a residual value. The non-zero saturation value observed in our numerical results is a finite-size effect due to the limited reservoir capacity; in the thermodynamic limit of an infinite bath, the entropy is expected to vanish entirely, consistent with global unitarity. In the lower panel, we plot the time evolution of the particle number within the system for different initial states and mark the instant at which half of the particles have leaked out with white open circles. When approximately half of the initial excitations have departed—corresponding to the Page time—the entanglement entropy attains its peak and transitions into a declining phase.
	
	
The bend down behavior of the Page curve, rather than saturation, ensures the unitarity of the entire process. If the bath is sufficiently large, all particles originally located in the system will gradually escape into the bath, eventually leaving behind an empty system in a pure state. A natural question then arises: an empty system contains no information, so where has the information originally stored in the system gone? How and when does this leakage occur? To further investigate this issue, we use this analog model to study the dynamical process of information transfer by computing the evolution of different quantum resources from the interior to the exterior.

	\paragraph{{\it Entanglement}.} In general relativity, the event horizon constitutes a causal boundary: under purely classical dynamics, any particle crossing it is permanently captured within the black hole. Quantum effects, however, modify this picture by rendering the horizon effectively permeable, leading to the emission of Hawking radiation. Although Hawking’s original semi-classical calculation described the radiation as purely thermal—and thus seemingly information free subsequent work indicates that a more complete quantum-gravitational treatment can introduce non-thermal features, such as those captured by the island rule or other quantum information preserving mechanisms. In this way, the radiation offers a potential avenue for resolving the information paradox, as interior information may gradually leak out through such corrected radiation or related quantum channels.
	
 	We now study how entanglement initially shared among particles inside the interior region (characterized by a negative coupling $\kappa<0$) is transmitted to exterior particles via the sequential emission of radiation quanta. It should be noted that the time evolution does not preserve unitarity for arbitrary subsystems. As a result, an initially pure state of a generic subsystem generally evolves into a mixed state. For instance, consider the reduced density matrix (RDM)$\rho _{AB}(0)$ of two qubits $A$ and $B$ in the external bath. Initially,
	\begin{equation}
		\operatorname{Tr}[\rho_{AB}(0)] = \operatorname{Tr}[\rho_{AB}(0)^2] = 1,
	\end{equation}
	whereas after an evolution time $\tau$ we typically find
	\begin{equation}
		\operatorname{Tr}[\rho_{AB}(\tau)^2] < 1.
	\end{equation}
	Since the resulting reduced density matrix is mixed, the von Neumann entropy obtained by tracing over one of the constituent qubits no longer serves as a faithful measure of the bipartite entanglement between $A$ and $B$. In such mixed-state configurations, $S_{\text{ent}}$ conflates quantum correlations with classical statistical uncertainties. 
	
	To isolate the purely quantum part of the correlations, we instead employ the $concurrence$, which provides a bound for the entanglement of formation in non-pure states. It is defined as
	\begin{equation}
		C(\rho_{AB}) = \max\bigl\{0,\; \lambda_1 - \lambda_2 - \lambda_3 - \lambda_4\bigr\},
	\end{equation}
	where \(\lambda_1 \geq \lambda_2 \geq \lambda_3 \geq \lambda_4 > 0\) are the square roots of the eigenvalues of the matrix \(\rho_{AB} \tilde{\rho}_{AB}\), with \(\tilde{\rho}_{AB} = (\sigma_y \otimes \sigma_y) \rho_{AB}^* (\sigma_y \otimes \sigma_y)\).
	We first consider qubit pairs with varying degrees of entanglement, initially prepared inside the system as described by Eq.~(\ref{eq_phi0ent}), while the external bath remains in its ground state. During the time evolution, as particles gradually escape from the interior into the bath, the entanglement initially stored inside the system is redistributed. The full system plus bath evolution follows the Liouville-von Neumann equation,  
	\begin{equation}
		\frac{d\rho_{\text{tot}}(t)}{dt} = \frac{i}{\hbar}[\rho_{\text{tot}}, H].
	\end{equation}  
	To investigate the entanglement dynamics within the external bath, we consider a bipartite subsystem composed of two specific qubits, labeled \(A\) and \(B\). The reduced density matrix (RDM) of this bipartite subsystem is obtained by tracing out the remaining degrees of freedom of the total system
	\begin{equation}
		\rho_{AB} = \operatorname{Tr}_{\overline{AB}} \rho_{\text{tot}},
	\end{equation}
	where $\operatorname{Tr}_{\overline{AB}}$ denotes the partial trace over the complement of the $\{A,B\}$ subsystem. We subsequently evaluate the temporal evolution of the concurrence between the two outermost qubits of the bath (\(A = L-1,\; B = L\)).
	\begin{figure}[t]
		\centering{\includegraphics[width=0.96\linewidth]{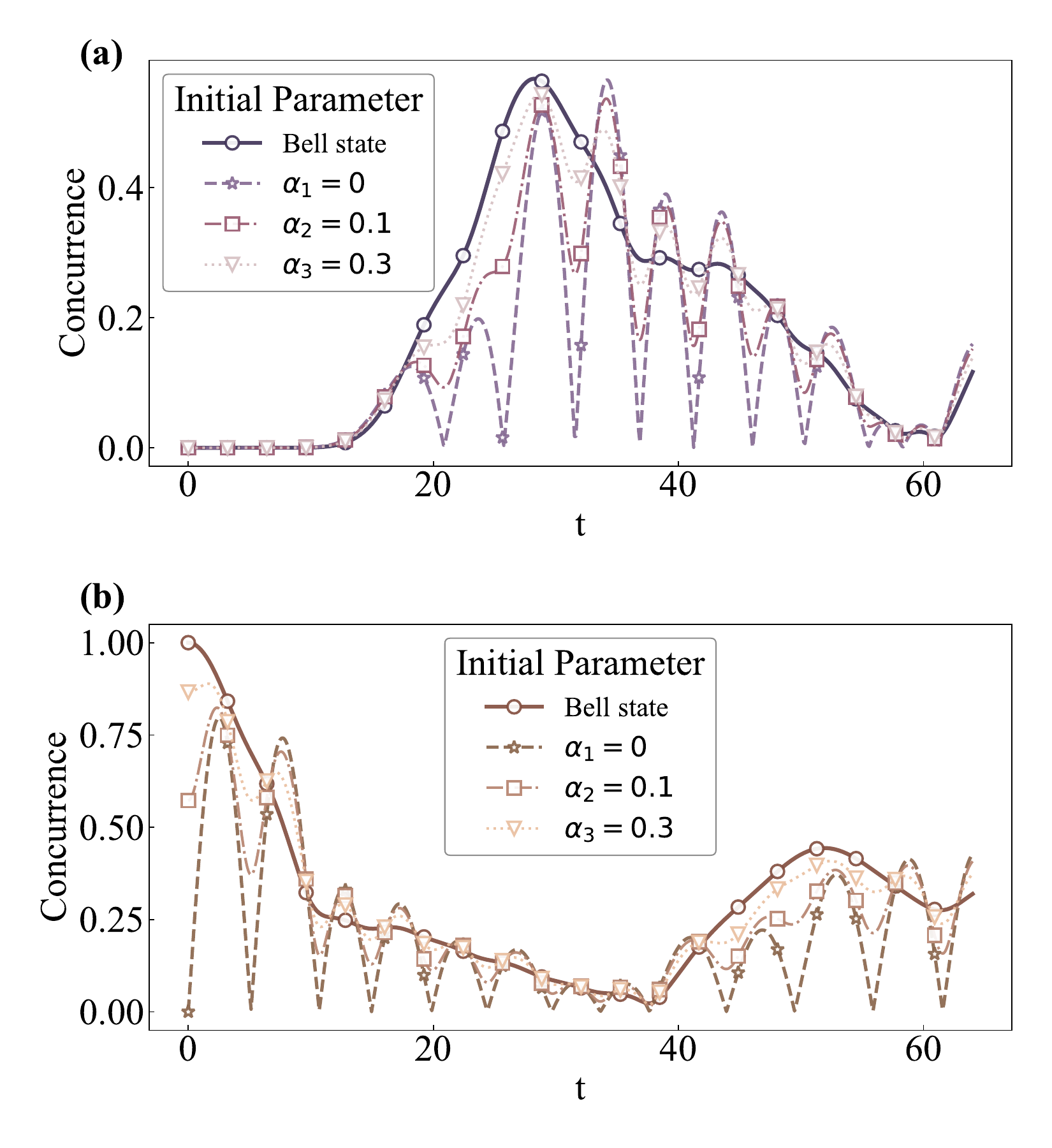}
			\caption{(a) Time evolution of the concurrence between the two outermost qubits in the bath (sites $L$-1 and $L$). (b) Time evolution of the concurrence between two nearest neighbor qubits inside the system (sites 1 and 2). In both panels, the first two qubits of the system are initialized with different degrees of entanglement, parameterized by \(\alpha\). The numerical data are obtained for a chain of length \(L=10\) and a discretization interval \(d=2\).}\label{fig_entang}
		}
	\end{figure}
	
 	Fig~\ref{fig_entang}(a) displays the results for four initial states with increasing internal entanglement: \(\alpha = 0\) (pentagon), \(\alpha = 0.1\) (square), \(\alpha = 0.3\) (inverted triangles), and \(\alpha = 1/\sqrt{2}\) corresponding to a maximally entangled Bell state (circles). At early times (\(t < 10\)), no entanglement is observed between \(A\) and \(B\). This is expected because the bath is initially empty, and entanglement can only emerge after interior particles have escaped into the bath. Although the XY interaction itself is local (coupling only nearest neighbor spins), the time evolution operator that generates the unitary evolution of the full chain is non-local. This effective non-locality facilitates the generation and propagation of quantum entanglement across the chain, allowing correlations to spread from the interior region toward the asymptotic bath. For the case $\alpha=0$, the time evolution commencing from an unentangled initial state yields oscillatory entanglement in the bath qubits, characterized by peaks that are invariably followed by substantial decay.
	
	When the system qubits harbor initial entanglement ($\alpha>0$), the dynamical behavior undergoes a qualitative transformation. Although bath qubit entanglement continues to display rapid, low amplitude oscillations akin to weak noise, it no longer attenuates to zero. With increasing initial system entanglement, the troughs in bath entanglement rise progressively, while the relative amplitude of these oscillations diminishes correspondingly. This indicates that the entanglement observed in the bath is not merely a transient effect of particles leaking out; its decay profile is also tied to the initial state inside the system. In other words, the entanglement originally stored in the interior system appears to leak into the bath, preventing the complete disappearance of bath entanglement.
	
	This behavior can be observed in Fig.~\ref{fig_entang}(b), which displays the entanglement dynamics inside the system. The entanglement exhibits rapid oscillations superimposed on an overall decaying trend, similar to the behavior observed in the external bath. For initial states with no intrinsic entanglement inside the system, the entanglement amplitude decays nearly to zero after each oscillation. As the initial entanglement within the system increases, the oscillatory component becomes less pronounced. When the internal qubit pair is initialized in a maximally entangled state, these oscillations are almost entirely suppressed, and the resulting curve effectively forms the envelope of the trajectories corresponding to other initial conditions.
	In the early stage of evolution ($t < 10$), particles escape rapidly from the system into the bath, which is initially empty. After $t > 10$, approximately half of the internal particles have already escaped, and the subsequent dynamics become more gradual. If the bath is sufficiently large, all particles originally inside the system will eventually flow out by the late stage of evolution, leaving the system empty. At this point, the entanglement inside the system completely decays to zero, corresponding to the final stage of total black hole evaporation.
	From the perspective of black hole information dynamics, these results suggest a broader implication: if the underlying evolution is unitary, then the particles radiated from the black hole interior may not be purely thermal. Instead, they can carry subtle quantum correlations that partially encode information initially stored within the black hole.
	
	\paragraph{{\it Coherence}.} Quantum coherence is a fundamental quantum mechanical property that lies at the heart of many distinct physical phenomena and is essential for a wide range of quantum tasks, such as quantum metrology, computation, and thermodynamics.
	\begin{figure}[t]
		\centering{\includegraphics[width=0.96\linewidth]{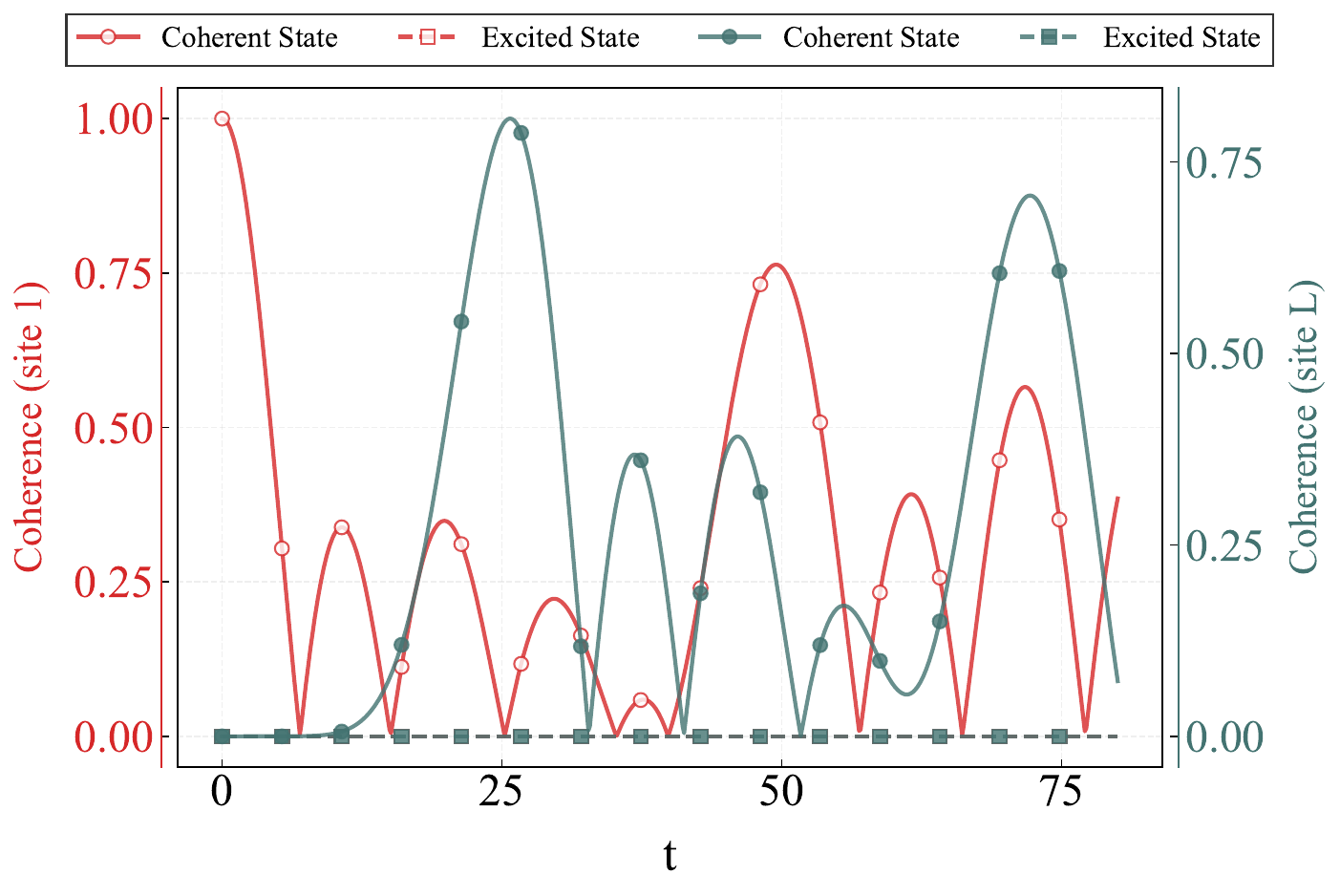}
			\caption{Time evolution of coherence, quantified via the $l_1$-norm, for two distinct qubits: the first qubit at site 1 within the system (red traces) and the final qubit at site $L$ in the bath (green traces). Two initial configurations for the first qubit are contrasted: the excited state (square markers) and the maximally coherent state (circular markers), with all other qubits initialized in the spin-down state. Numerical results are obtained for a chain of length $L=10$ and discretization interval $d=2$.}\label{fig_coher}
		}
	\end{figure}
	In recent years, coherence has attracted considerable theoretical and experimental interest within quantum information science~\cite{Monz:2011znf,Streltsov:2015xia,Engel:2007zzb,Xiao:2025flt,Qin:2025npi,Petkova:2025cil,Shi:2022elq}. More formally, under the framework of resource theory, it has been systematically characterized as a quantifiable and operational quantum resource. This theory establishes that coherence can serve as a universal ``fuel'': by consuming a maximally coherent state alongside free incoherent operations, one can in principle realize any other coherent operation~\cite{Baumgratz:2013ecx,Streltsov:2016iow,Hu:2018mni,Chitambar:2018rnj}.

	Within the resource-theoretic formulation, one first fixes a reference basis \(\{\ket{i}\}\). The set of incoherent states is then defined as all density matrices that are diagonal in this basis, i.e., states of the form~\cite{Baumgratz:2013ecx}
	\begin{equation}
		\hat{\delta} = \sum_{i=1}^{d} \delta_i \ket{i}\bra{i}, 
	\end{equation}
	where \(\delta_i \geq 0\) and \(\sum_i \delta_i = 1\). These states form the set of free states that possess no coherence,
	while the maximally coherent state is defined as~\cite{Baumgratz:2013ecx}
	\begin{equation}
		\ket{\phi} := \frac{1}{\sqrt{d}} \sum_{i=1}^d \ket{i}.
	\end{equation}
	For a qubit, selecting the spin-up $\ket{\uparrow}$ and spin-down $\ket{\downarrow}$ states as the basis, then the maximally coherent state can be defined as 
	\begin{equation}
		\ket{+} = \frac{1}{\sqrt{2}} (\ket{\uparrow} + \ket{\downarrow}),
	\end{equation} 
	and the free states take the form
	\begin{equation}
		\hat{\rho}_{free} =  c_i \ket{\uparrow}\bra{\uparrow}+(1-c_i) \ket{\downarrow}\bra{\downarrow}. \label{eq_free}
	\end{equation}
	To elucidate coherent dynamics in this spin-chain model with an effective horizon, the initial state of the first qubit within the system is prepared in one of two configurations: the excited state $\ket{\phi_e} = \ket{\uparrow}$ or the maximally coherent state $\ket{\phi_c} = \ket{+}$. The coherence emergence in the final bath qubit is quantified using the widely adopted $l_1$-norm measure~\cite{Baumgratz:2013ecx}
	\begin{equation}
		C_{l_1}(\rho) = \sum_{i \neq j} |\rho_{i,j}| =|\bra{\uparrow}\rho\ket{\downarrow}|+|\bra{\downarrow}\rho\ket{\uparrow}|,
	\end{equation}  
	revealing distinct leakage behaviors. 
	
	As illustrated in Fig.~\ref{fig_coher}, two distinct initial configurations for the first system qubit are examined: the excited state $\ket{\phi_e} = \ket{\uparrow}$ (denoted by square markers) and the maximally coherent state $\ket{\phi_c} = \ket{+}$ (denoted by circle markers), with all other qubits initialized in the ground state $\ket{\phi_g} = \ket{\downarrow}$. Coherence is measured for site 1 in the system (red traces) and site $L$ in the bath (green traces). For the initial excited state $\ket{\phi_e}$ (squares), neither the system nor the bath harbors initial coherence, and none emerges or propagates to the probed qubits during the evolution.  
	In stark contrast, the initial maximally coherent state $\ket{\phi_c}$ (circles) yields rapid decay of system coherence in the early stages; following a transient period, as system particles egress to the bath, coherence diffuses to the bath qubit, manifesting a swift upsurge followed by pronounced temporal oscillations. This indicates that no new coherence is generated during the spin-chain evolution; rather, the pre-existing coherence within the initial system is progressively transferred into the bath along with the leaking particles. During the early stage, a fraction of the particles and coherence is transferred to the external bath, causing the internal subsystems to exhibit mixed-state characteristics. This, in turn, drives a continuous increase in the entanglement entropy between the initial system and the bath. Subsequently, as a substantial amount of particles and coherence is transferred to the external bath, the internal system undergoes a continuous purification toward the empty state, and the entanglement entropy declines steadily.
	\begin{figure}[t]
		\centering{\includegraphics[width=0.96\linewidth]{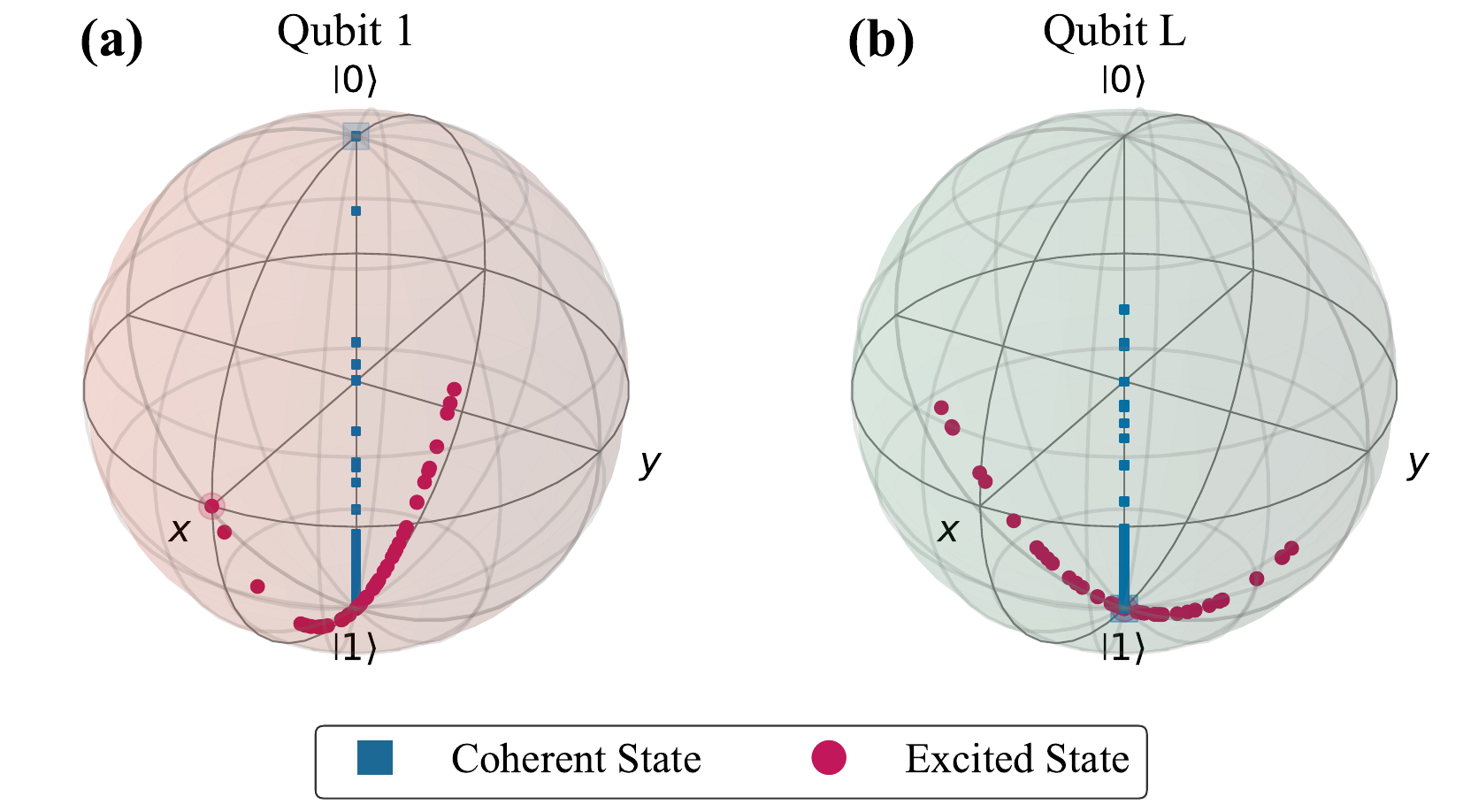}
			\caption{Trajectories of quantum state evolution on the Bloch sphere. Panels (a) and (b) depict measurements on distinct qubits: site 1 in the system for (a) and site $L$ in the bath for (b). Two initial states for the first system particle are considered: the spin-up excitation $\ket{\uparrow}$ (marked by blue squares) and the coherent superposition $\ket{+}$ (marked by pink circles). The points on the Bloch sphere corresponding to these initial states are enlarged with the respective markers. Numerical results are obtained for a chain of length $L = 10$ and discretization interval $d = 2$.}\label{fig_bloch}
		}
	\end{figure}
	
	To further elucidate the coherence dynamics throughout the evolution, we present the Bloch-sphere evolutions for the aforementioned initial configurations and monitored qubits. The position $\boldsymbol{r}$ of the quantum state on the Bloch sphere obeys  
	\begin{equation}		
		\rho = \frac{I + \boldsymbol{r} \cdot \boldsymbol{\sigma}}{2}, \quad \boldsymbol{\sigma} = (\sigma^x, \sigma^y, \sigma^z),
	\end{equation}  
	where $\boldsymbol{r} = \langle \sigma^x \rangle, \langle \sigma^y \rangle, \langle \sigma^z \rangle$. In the basis $\{ \ket{\uparrow}, \ket{\downarrow} \}$, eq.~(\ref{eq_free}) implies that any free state resides along the $z$-axis of the Bloch sphere.  
	
	As shown in Fig.~\ref{fig_bloch}, trajectories are depicted for distinct qubits: site 1 in the system for (a) and site $L$ in the bath for (b). Initial states comprise the coherent state $\ket{+}$ (square markers) and the excited state $\ket{\uparrow}$ (circle markers), with each trajectory's initial point enlarged via the corresponding marker. For the initial excited state, the quantum-state trajectory oscillates strictly along the $z$-axis, signifying that this process generates no coherence. In contrast, for the initial coherent state, the first qubit rapidly evolves toward the free state \(\ket{\downarrow}\), whereas the external qubit remains close to \(\ket{\downarrow}\) for a prolonged period (appearing as a tight cluster on the Bloch sphere) before departing abruptly from the set of free states. This pattern signifies that coherence is transported by leaking particles rather than engendered by the system's intrinsic dynamics; accordingly, progressive particle leakage to the bath concomitantly depletes the coherence reservoir.

{\bf Discussion}

	In this work, we construct an effective horizon using a one-dimensional inhomogeneous XY spin chain, thereby simulating the behavior of quantum fields in curved spacetime. Within a system-plus-bath framework, the model exhibits Page-curve-like entanglement entropy dynamics. We further investigate how the initial entanglement and coherence within the system progressively escape through the effective horizon. In our model, the quantum field equations in curved spacetime are mapped onto the Heisenberg equations of the spin chain via a rigorous coordinate discretization~\cite{Yang:2019kbb,Liu:2024wqj}. In this construction, the coupling $\kappa$ near the effective horizon vanishes only when the discretization interval approaches zero. Consequently, propagation in both directions is suppressed at the horizon, a feature consistent with the perspective of an observer in the corresponding coordinate system. Reference~\cite{Liu:2024wqj} discusses how particle evolution depends on the coordinate systems of different observers within an analogue black hole background. Moreover, with a sufficiently small discretization interval and an adequately large system size, this scheme can characterize the 'trapping' of particles inside the horizon and capture the thermal spectrum of the radiation. 

\begin{figure}[t]
		\centering{\includegraphics[width=0.96\linewidth]{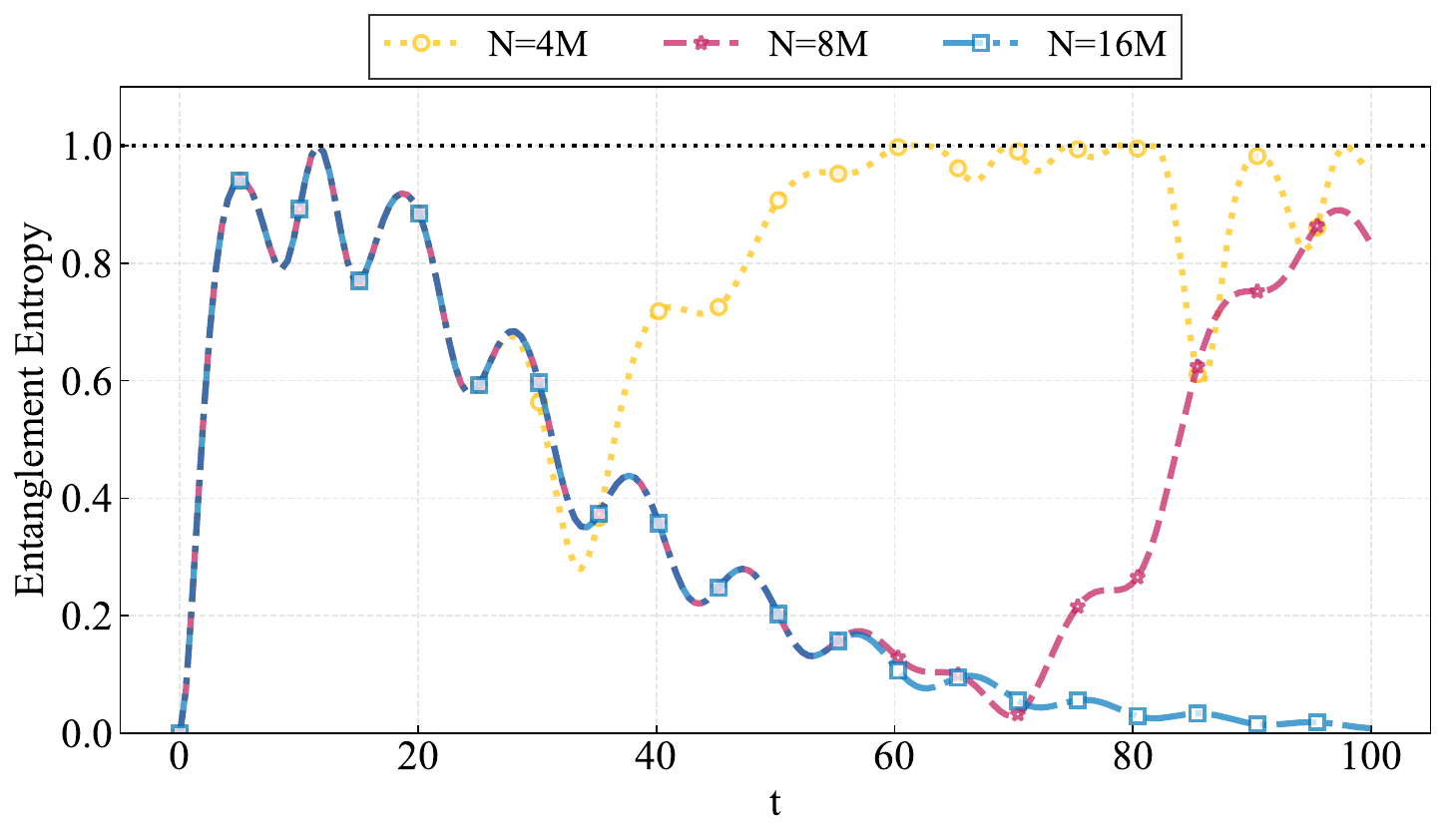}
			\caption{Size dependence of entanglement entropy evolution in the single-excitation subspace. The system size is fixed at $M=2$, while the bath size is varied as $N=4M$ (yellow circles), $8M$ (pink stars), and $16M$ (blue squares). The initial state is prepared with a single excitation localized at site $M$, with a discretization interval of \(d=2\). }\label{fig_size}
		}
	\end{figure}

	While several studies have constructed effective horizon geometries at the field-theory level to demonstrate distinct advantages~\cite{Kinoshita:2024ahu,Konye:2022wds,Kinoshita:2025fij,Horner:2022sei}, the inhomogeneous-chain construction adopted here offers a unique trade-off: a moderately small discretization interval can effectively enhance the particle tunneling rate, enabling us to capture substantial particle radiation and access the intermediate-to-late stages of evaporation. Consequently, the Page-curve-like dynamics of the entanglement entropy can be resolved. It should be noted that an excessively fine discrete interval leads to weak coupling coefficients, which in turn requires longer evolution times and larger external scales to ensure substantial tunneling of interior particles, accompanied by oscillations arising from interior finite-size effects. Nevertheless, the overall trend of the Page curve remains unchanged. To further investigate the scalability of our findings, we examine the effect of an extended external bath on the entanglement entropy dynamics within the single-excitation subspace, as shown in Fig.~\ref{fig_size}. In these simulations, the system size is fixed at $M=2$. We vary the bath size $N$ as $4M$, $8M$, and $16M$. The case $N=4M$ (yellow circles) corresponds to the default configuration used in the main text. The other cases, $N=8M$ and $16M$, are represented by pink stars and blue squares, respectively. We observe that the late-time residual entanglement entropy decays rapidly as the bath size increases; specifically, when the bath size exceeds $8M$, the residual value asymptotically approaches zero. This behavior aligns with our analysis of Fig.~\ref{fig_pagecur} and is consistent with the conclusions reported in Ref.~\cite{Kehrein:2023yeu}. Furthermore, this horizon construction is well-suited for implementation on existing superconducting quantum experimental platforms~\cite{Shi:2021nkx}. It should be noted, however, that while our model captures Page-curve-like dynamics, the emission process does not strictly exhibit a thermal spectrum. This is anticipated under globally unitary evolution: as approximately half of the excitations leak into the bath, the system begins a continuous purification process until it reaches an empty state, at which point the bath eventually returns to a pure state.
	
{\bf Conclusion}

	In summary, we have investigated the quantum information dynamics of a spin-chain model with an effective horizon realized via an isotropic, site-dependent XY spin chain. Our model successfully demonstrates Page curve like behavior, characterizing the evolution of entanglement entropy between the black hole interior and the external bath. By analyzing the temporal decay of local occupation numbers, we identified a clear correspondence between the depletion of interior excitations and the ``Page time," marking the onset of information recovery.  Furthermore, by monitoring bipartite correlations within the external bath, we evaluated the transport of initial entanglement and coherence from the interior to the asymptotic region. Our results reveal that the dynamics of this transfer are highly sensitive to the initial state configuration. While low-entanglement initializations result in rapid transient oscillations of the transmitted concurrence, we find that maximally entangled Bell states facilitate a significantly more continuous and stable transport process. These findings provide a robust quantum simulation framework for exploring the evaporation of black holes in curved spacetimes, offering critical insights into the interplay between many-body physics and the resolution of the black hole information paradox.
 \\
 
{\bf Acknowledgements}  

    This work was supported by the National Natural Science Foundation of China under Grants No. 12475051, and No. 12421005; the science and technology innovation Program of Hunan Province under grant No. 2024RC1050;  the innovative research group of Hunan Province under Grant No. 2024JJ1006; the Hangzhou Leading Youth Innovation and Entrepreneurship Team project  under Grant No. TD2024005; and the scientific research start-up funds of Hangzhou Normal University: 4245C50224204016.

%

\end{document}